\documentclass[12pt]{article}
\usepackage{epsfig}
\def\qt{\rm{Q_{tot}}}
\def\qs{\rm{Q_{sat}}}
\def\Ybar{\rm{\bar{Y}}}
\def\2n{\rm{2^n}}
\def\lsat{\rm{L_{sat}}}
\def\ltest{\rm{L_{test}}}
\def\tbar{\rm{< t >}}
\def\ddt1{\rm{\delta t_{-}}}
\def\dt2{\rm{\delta t_{+}}}

\def\dtnosat{\rm{\delta t_{unsat}}}

\begin{document}
\hfill AS-TEXONO/03-03 \\
\hspace*{1cm} \hfill \today

\begin{center}
%%\vskip 0.5cm
{\Large  \bf
Effective Dynamic Range in Measurements\\
with Flash Analog-to-Digital Convertor
}\\
\vskip 0.5cm
\large
Q.~Yue$^{a,b}$,
W.P.~Lai$^{c,d}$,
W.C.~Chang$^{c}$,
H.B.~Li$^{c}$,
J.~Li$^{a,b,c}$,\\
S.T.~Lin$^{c}$,
D.Z.~Liu$^{b}$,
J.F.~Qiu$^{a}$,
V.~Singh$^{c}$,
H.T.~Wong$^{c,}$\footnote{Corresponding~author:
Email:~htwong@phys.sinica.edu.tw;
Tel:+886-2-2789-6789;
FAX:+886-2-2788-9828.}

\normalsize
\vskip .2cm

\begin{flushleft}
{$^{a}$\rm 
Institute of High Energy Physics, Beijing 100039, China.\\}
{$^{b}$\rm 
Department of Engineering Physics, Tsing Hua University, 
Beijing 100084, China\\}
{$^{c}$\rm
Institute of Physics, Academia Sinica, Taipei 115, Taiwan.\\}
{$^{d}$\rm 
Department of Management Information Systems,
Chung Kuo Institute of Technology, \\
\hspace*{1cm} Hsin-Chu 303, Taiwan.\\}
\end{flushleft}

\end{center}
\vskip 0.5cm
\begin{abstract}

Flash Analog to Digital Convertor (FADC)
is frequently used in
nuclear and particle physics
experiments, often 
as the major component in  
big multi-channel systems.
The large data volume makes
the  optimization of operating parameters necessary.
This article reports a study of a method
to extend the dynamic range
of an 8-bit FADC from the nominal $\rm{2^8}$ value.
By comparing the integrated pulse area 
with that of a reference profile, 
good energy reconstruction and event identification
can be achieved on saturated events from
CsI(Tl) crystal scintillators.
The effective dynamic range can be
extended by at least 4 more bits.
The algorithm is generic and is expected to
be applicable to other detector systems
with FADC readout. 

\end{abstract}

\begin{flushleft}
\end{flushleft}

\vfill

\begin{center}
{\it ( Submitted to Nucl. Instrum. Methods A ) }
\end{center}

\newpage

\section{Introduction}

With the advance of electronics, data acquisition (DAQ)
and computing capabilities,
it is more and more common to use
Flash Analog to Digital Convertors (FADCs)
as the general-purpose
readout device for different detectors.
Typically, the entire pulse shape from
the individual detector module,
as well as the relative
timing between pulses from different 
detector channels, are recorded.
Such readout scheme replaces
Analog Digital Convertors
and Time-to-Digital Convertors,
which give information only on the
amplitude or integrated area and the timing of the
pulses, respectively.

With FADCs as readout,
the information related to a single
pulse is represented by many measurements.
Frequently, multi-channel (order of hundreds
and thousands is common) FADC systems
are deployed. As a result, a large volume
of data have to be extracted, stored and processed. 
The {\it n}-bit resolution per time-bin
and the sampling rate of the FADC  
are important parameters
which  will have tremendous effect
on the characterization and
performance of the experiment
such as detector resolution,
DAQ dead time, 
data size, data processing
time, cost and so on.

A motivated question to address
is therefore ``What is the dynamic range of 
an n-bit FADC, beyond the nominal
$\rm{2^n}$ value?''
That is, can the crucial information
like the total area as well as
the rise and fall times
be reconstructed, even though
part of the pulse $-$
the largest and usually the most important part $-$
is not recorded?
These are the focuses of investigation 
of this article.

In Section~\ref{sect::basic},
we describe the research program on which this
project is based, and the hardware
used to generate the data which allow
such an investigation.
Two approaches are surveyed and compared.
One of the algorithms,
the Charge Summation Correction Method,
showed better potentials and were pursued
in depth. The goals are to achieve
both energy reconstruction and particle
identification even for saturated events.
The details and results are discussed in 
Section~\ref{sect::partq}.

Though the algorithm was developed and tested
with scintillating CsI(Tl) crystal detectors,
the approach is generic and can be applicable 
to other detector schemes using FADC
readout as well.

\section{Scintillating CsI(Tl) Crystal Detector}
\label{sect::basic}

The present investigation is part
of the research program of the
TEXONO Collaboration~\cite{ksexpt}.
A CsI(Tl) scintillating crystal~\cite{proto} 
array with 2~kg modular mass and a
total mass of 186~kg is taking data
at the Kuo-Sheng (KS) Nuclear Power Plant
to study low energy neutrino physics.
Each module is read out by two 
photo-multipliers (PMTs) and the 
signals are recorded by a 20~MHz
FADCs with 8-bit resolution~\cite{electronics}. 
The FADC value at the $\rm{t_i^{th}}$ time-bin
is denoted by $\rm{y_i}$. 
Throughout this work, 
we use the normalized value $\rm{Y_i}$ which is
$\rm{y_i}$ divided by the
saturation level $\lsat = \2n - 1 = 255$ ~~:
$\rm{ Y_i =  y_i / \lsat }$ ~ .

The physics requirements 
of the experiment are that events
with energy from 10~keV to 10~MeV can be
reconstructed, and that pulse
shape discrimination (PSD) can be
applied to these events to achieve
event ($\alpha$/$\gamma$) identification.
That is, a dynamic range of at least 1000
is necessary with an 8-bit resolution
readout per time-bin.

Special data were taken to devise techniques
and to develop software algorithms addressing 
this issue.
A $\rm{5 \times 5 \times 5 ~ cm^3}$
CsI(Tl) crystal was used
and signals were read out by
a single PMT
and a FADC channel at
identical settings to the KS experiment.
The trigger was set at the $\rm{t_0 = 100^{th}}$ 
FADC time-bin, such that
the pulse information of $\ddt1$=5~$\mu$s (100 time-bins)
and $\dt2$=25.6$~\mu$s (512 time-bins)
before and after the trigger
were recorded, respectively.
Both $\alpha$ and $\gamma$ events were
taken by activating the CsI(Tl) crystal
with $^{241}$Am-Be and $^{137}$Cs sources, respectively.

\begin{figure}
\epsfig{file=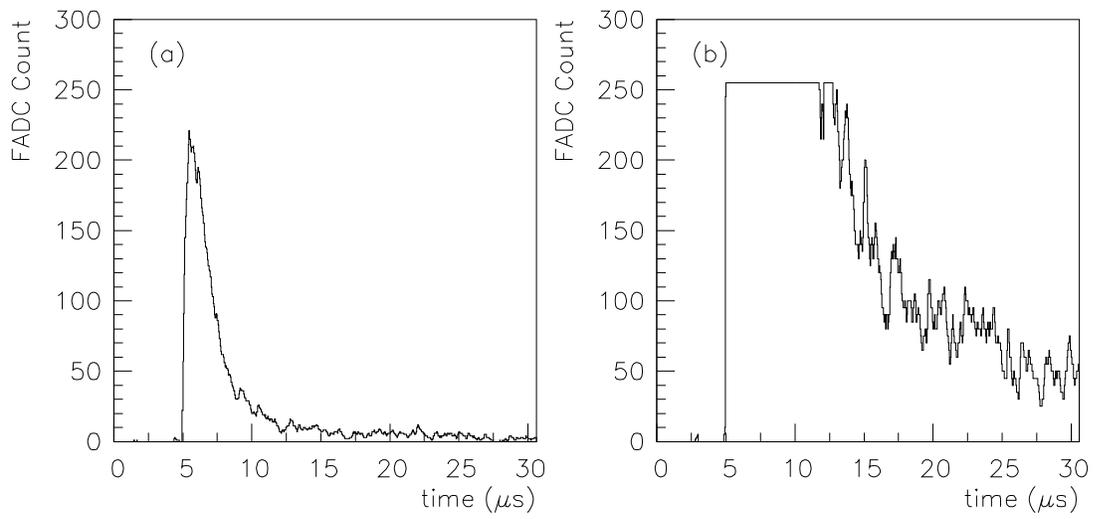,width=15cm}
\caption{
Typical (a) unsaturated and (b) saturated pulses recorded 
by the FADC.
}
\label{singlepulse}
\end{figure}

The amplitude of the pulse was adjusted by
selecting the appropriate PMT high voltage.
Typical unsaturated  and saturated pulses
are displayed in Figures~\ref{singlepulse}a and
\ref{singlepulse}b, respectively.
A large sample of unsaturated events
are added together to produce
a smooth reference pulse shape $\Ybar$(t),
where the peak amplitude is normalized to 1.
The profiles of $\Ybar$(t) due to $\alpha$- and $\gamma$-events
are depicted in Figure~\ref{pulseshape}, 
and can be fitted to
the function~\cite{proto}
\begin{equation}
\label{eq::shape}
\rm{
\Ybar (t) ~ =  N \ast [ ~ 1 - exp ( - \frac{t}{\tau_0} ) ~  ]
\ast
[ ~ \frac{1}{\tau_1} ~ exp ( - \frac{t}{\tau_1} )
+ \frac{r}{\tau_2} ~ exp ( - \frac{t}{\tau_2} ) ~ ] ~ . 
}
\end{equation}
The rise time($\rm{\tau_0}$)
and fall times($\rm{\tau_1 , \tau_2}$)
as well as the ratio between the slow and fast decay
components(r) are listed in Table~\ref{time}.
The normalization constant N is chosen
to make the peak amplitude equal to unity.
The values of $\rm{\tau_0}$ are dominated by
the electronics shaping rise time
of 250~ns~\cite{electronics},
while ($\rm{\tau_1 , \tau_2}$)
are due to the characteristic fluorescence
time profile of CsI(Tl).

The integrated charge (Q) of an event is obtained by
numerically integrating the 
pedestal-subtracted FADC pulses:
\begin{equation}
\rm{
Q ~ = ~ \sum_{t_i > t_0} ~ [ ~ Y_i  - Y_0 ~ ] ~  = 
~ \sum_{t_i > t_0}  ~  { [ ~  y_i - y_0 ~  ] \over \lsat  } ~~~ ,
}
\end{equation}
where $\rm{y_0}$ is the average pedestal level 
determined from the pre-trigger part 
($\rm{t_i < t_0}$) of the pulse,
and $\rm{Y_0 = y_0 / \lsat}$.
The typical full-width-half-maximum (FWHM)
resolution for unsaturated events at 662~keV
is about 8\%.

\section{Charge Summation Method}
\label{sect::partq}

\subsection{Initial Survey}
\label{sect::survey}

Saturated events are those with pulse
amplitude larger than those corresponding
to the maximum FADC value of  $\lsat = 255$.
The focus of this section is to devise
an algorithm with which energy
reconstruction and event identification
by PSD  can be achieved for these
class of events.

Two approaches were surveyed as
initial studies:
\begin{enumerate}
\item {\bf Statistical Approach} 
In principle, one can derive 
the values of N in Equation~\ref{eq::shape},
using maximum likelihood analysis
or minimum $\chi ^2$ fitting methods,
based only on the ($t_i$,$\rm{y_i}$)
data  where $\rm {y_i < \lsat}$.
The values of $\rm{( \tau_0 , \tau_1 , \tau_2 , r )}$
are adopted from those in Table~\ref{time}.
In practice, the fits are computer-time
consuming and unstable. 
The efficiencies are bad, especially when
the saturation is severe.
The main reason is that the
values of $\rm{\tau_0}$ and $\rm{\tau_1}$
are derived from small portions of the 
pulse shape. When the data from
these portions are missing due to saturation,
there will be insufficient constraints such that
it is difficult for the analysis procedures
to converge to the optimal values for N.
\item {\bf Pulse Analysis Approach}
The objective of this approach is to identify
measure-able values which are 
related to the physics variables
of interest, namely, the total
energy and the decay time.
The observables  are the
``saturated charge'' and
``mean time''.
This method uses all
the available information from the saturated
pulses.
Such an approach can be compared to
$-$ and in fact inspired by $-$ 
the ``Double Charge Method''~\cite{psddc} 
commonly-used to achieve $\alpha$/$\gamma$
identification in liquid~\cite{liqscin} and 
crystal~\cite{proto} scintillators.
\end{enumerate}

Results of the initial surveys
indicated that the pulse analysis approach 
would show better promises.
The ``Charge Summation Method''
based on this approach was therefore
pursued in details.
We discuss the algorithm, its consistency
checks and performance in energy
reconstruction and event identification
in the following sub-sections.

\subsection{Reference Profile and Correction Function}

From a standard reference pulse shape,
a relationship is obtained between
the ``saturated charge ($\qs$)'' defined by the
integration of the saturated events
and the ``total charge ($\qt$)'' from that of the 
raw fully-recorded pulses. 

\begin{figure}
\epsfig{file=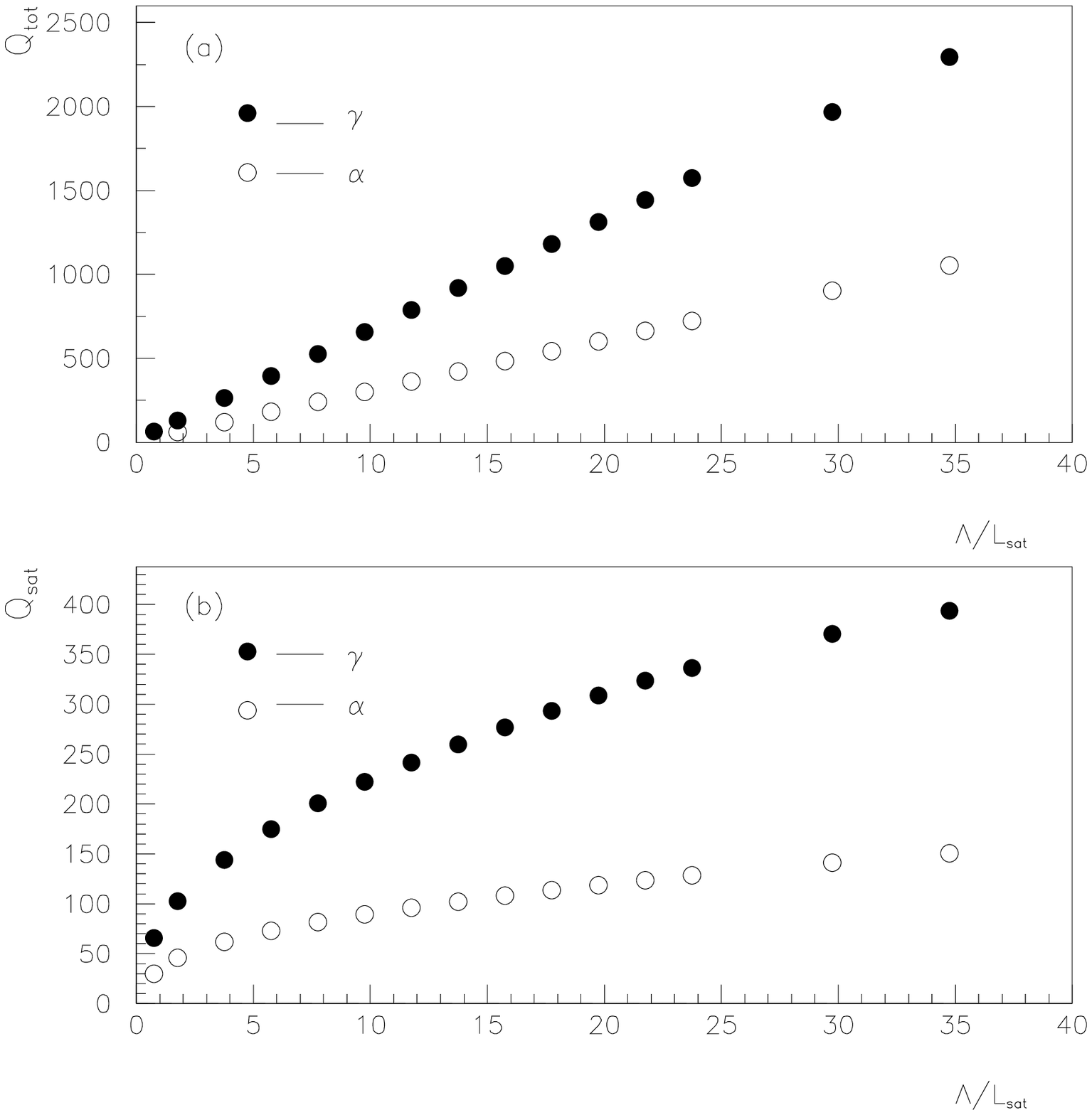,width=15cm}
\caption{ The dependence of (a) the total
charge $\qt$ and (b) the saturated charge $\qs$
on $\Lambda / \lsat$.
}
\label{qtqsamp}
\end{figure}

The saturation level for an n=8-bit FADC
is $\lsat = \2n - 1 = 255$.
To investigate the saturation effects on a 
pulse shape
with amplitude $\Lambda > \lsat$,
a profile $\rm{Z ( \Lambda )}$
is constructed from the reference
shape $\rm{\Ybar (t)}$ 
of Figure~\ref{pulseshape},
scaled-up by a factor of of $\Lambda$/$\lsat$.
Therefore, the value at the $\rm{t_i^{th}}$ time-bin is
given by
\begin{equation}
\label{eq::beta}
\rm{
Z_i ( \Lambda ) ~ = ~ \Ybar _i ~  [ ~ \frac{ \Lambda }{\lsat} ~ ] ~~ .
}
\end{equation}
The pedestal level $\rm{Z_0}$ is
the average of $\rm{Z_i}$ for $\rm{t_i<t_0}$.
The  dependence of the 
total and saturated charge as functions of $\Lambda$
for $\gamma$-events can be defined as:
\begin{equation}
\rm{
{\qt} (\Lambda) ~ = ~ \sum_{t_i} ~ [ ~ Z_i  - Z_0 ~ ]
}
\end{equation}
and
\begin{equation}
\rm{
{\qs} (\Lambda) ~ = ~ \sum_{t_i} ~ [ ~ \xi _i - Z_0 ~ ] ~,  
~~ where  ~~
\cases{
\rm{ ~ \xi _i ~ = ~ Z_i ~~~ for ~~~ Z_i  <  1 } \cr
\rm{ ~ \xi _i ~ = ~ 1 ~~~~  for ~~~ Z_i  \geq  1 } \cr}  ~~ ,
}
\end{equation}
respectively.
That is, given the profile $\rm{Z ( \Lambda )}$
with peak amplitude $\Lambda$/$\lsat$, $\qt$ and $\qs$ are
the integrated areas of the complete pulse
and the pulse truncated at a maximum amplitude of 
$\rm{ \Lambda = \lsat }$, respectively.
By varying the values of $\Lambda  / \lsat$, 
the dependence of  $\qt$ and $\qs$
on $\Lambda / \lsat$ for $\gamma$- and $\alpha$-events
can be derived.
They are depicted in Figure~\ref{qtqsamp}a and 
~\ref{qtqsamp}b, respectively.
As expected,
the variations shows a linear relationship
between $\qt$ and $\Lambda  / \lsat$,
while $\qs$ would level off at large $\Lambda  / \lsat$
due to saturation.

\begin{figure}
\epsfig{file=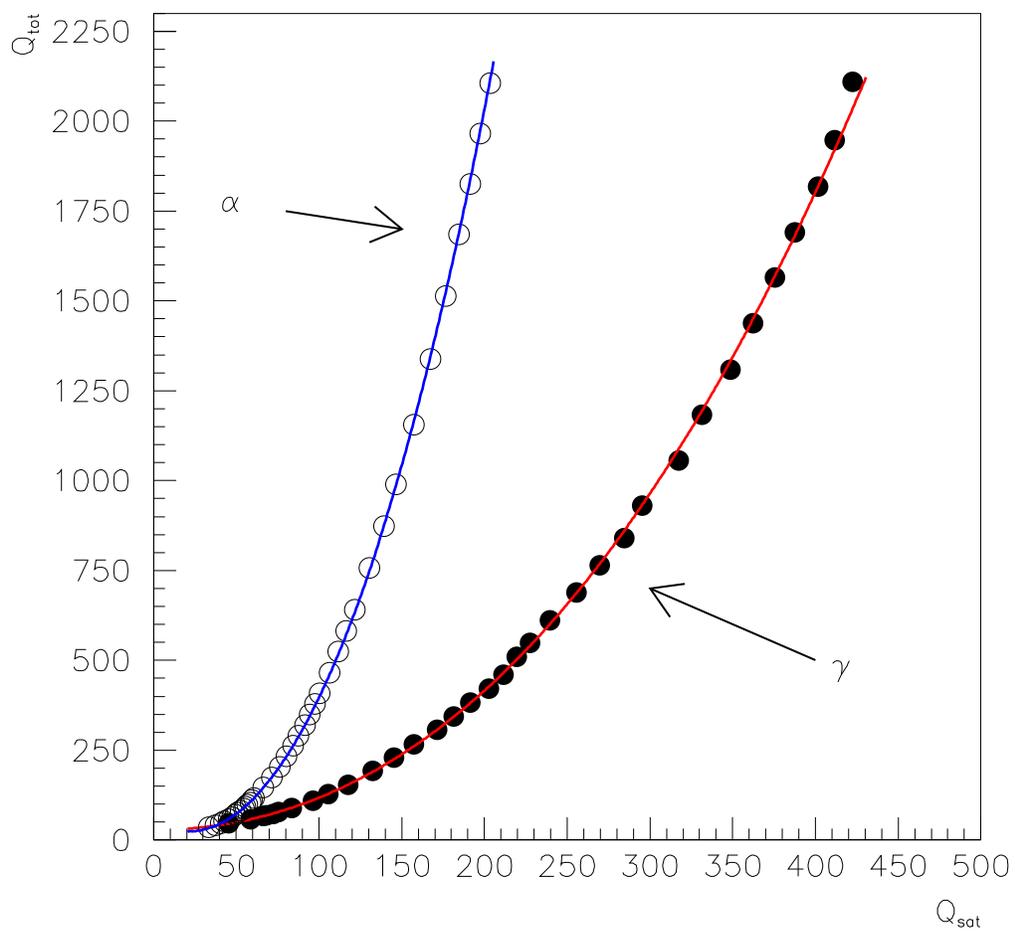,width=15cm}
\caption{ The total 
charge $\qt$ as a function of the saturated charge $\qs$, 
for both $\gamma$ and $\alpha$ events.  
The data are fitted to a polynomial function
for parametrization.
}
\label{qtqs}
\end{figure}

A relationship between $\qt$ and $\qs$
for the saturated events 
(that is, when $\qt > \qs$)
can then be obtained.
It is described by a polynomial function
\begin{equation}
\label{eq::qtqs}
\rm{
{\qt} ~ = ~ \sum_{k=0}^{3} ~ \alpha_k ~ {\qs} ^k
}
\end{equation}
for both the $\gamma$- and $\alpha$-samples
as displayed in Figure~\ref{qtqs}.
The coefficients $\rm{\alpha_i}$
are obtained from minimum $\chi ^2$-fits.

This function is then used to compensate the
missing area in saturated events.
Integration of the saturated pulses provide the
values of $\qs$, from which the
correct values of the  total charge $\qt$ can be derived in 
an event-by-event basis.

\subsection{Consistency Tests}

The correction algorithm discussed in the previous
section does not depend on the exact saturation
level $\lsat$. One can use this feature to 
perform consistency checks on data sample
where the total charge is known.
An unsaturated event is {\it made saturated} by software.
The algorithm is applied to the simulated saturated
event, and the corrected total charge is compared
with the raw measurements.

\begin{figure}
\epsfig{file=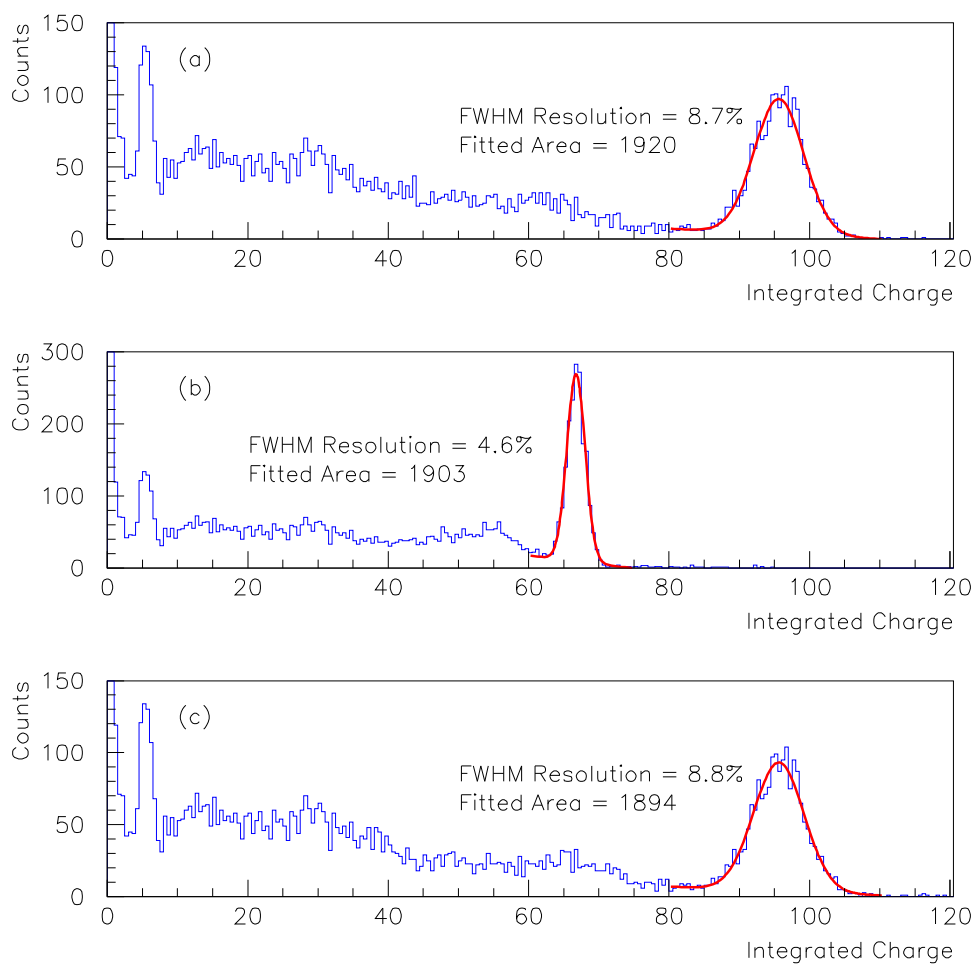,width=15cm}
\caption{ 
Energy Spectra for $^{137}$Cs:
(a) Raw spectra of unsaturated events,
(b) Spectra of events made saturated by software, and
(c) Spectra  of saturated events
after the corrections are applied.
}
\label{Xchecks}
\end{figure}

\begin{figure}
\epsfig{file=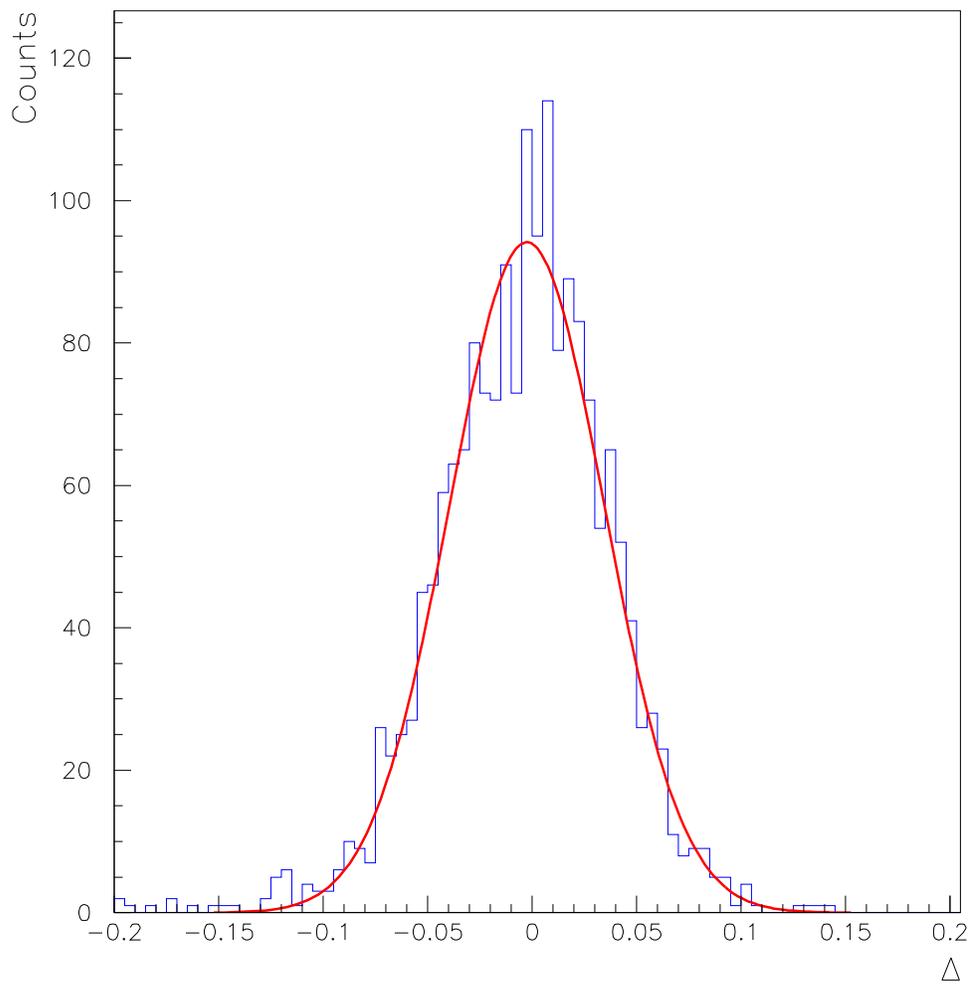,width=15cm}
\caption{
The deviation of 
the charge measurements between the
raw unsaturated events and 
the saturated events with corrections
applied.
}
\label{Deltaplot}
\end{figure}

A level of $\ltest$=100 is chosen. 
A new set of $\alpha _i$ 
is obtained by going through Eqs.~\ref{eq::beta}
to \ref{eq::qtqs} again, with $\lsat$ replaced
by $\ltest$.
A data set of $\gamma$-events was taken
with a $^{137}$Cs source.
The PMT high-voltage was adjusted such that
the amplitude for the 662~keV line is below
the actual saturation level of 255 but significantly
above  $\ltest$. 
The charge $\rm{Q}$ is measured by integrating 
the normalized FADC
counts 
$\rm{Y_i}$,
and  the spectrum is displayed in
Figure~\ref{Xchecks}a.
The same data is then made saturated by
re-defining $\rm{ Y_i \rightarrow Y^{\prime}_i }$
such that
\begin{equation}
\rm{
Y^{\prime}_i ~ = 
\cases{
\rm{ ~  Y_i ~~~~~~~~~~~~~~~  for ~~~~~~~ Y_i \leq \ltest / \lsat } \cr
\rm{ ~  \ltest / \lsat ~~~~~~  for ~~~~~~~ Y_i > \ltest / \lsat } \cr
} ~~ .
}
\end{equation}
The saturated charge is derived by integrating 
the normalized $\rm{Y_i^{\prime}}$
and the spectra is shown in Figure~\ref{Xchecks}b,
while that for
the corrected charge $\rm{Q^{\prime}}$ after the correction 
procedures is shown in Figure~\ref{Xchecks}c.
There is excellent agreement on the $^{137}$Cs peaks
in terms of their integrated charge, 
resolution and area  between 
Figures~\ref{Xchecks}a and \ref{Xchecks}c.
The event-by-event deviations 
are parametrized by
\begin{equation}
\rm
{
\Delta ~ = ~ { Q - Q^{\prime} \over ( ~ Q + Q^{\prime} ~ ) /2}  ~~~~~ .
}
\end{equation}
The distribution of $\Delta$
is depicted in Figure~\ref{Deltaplot}.
The results show that 
the mean bias is less than 0.5\% 
which is small compared to
the intrinsic FWHM resolution of 8.7\%. 
Therefore, the correction procedures are valid
and justified.

\subsection{Results}

\subsubsection{Energy Reconstruction}

The algorithms are applied to actual saturated data.
The data is generated by increasing the PMT high voltages
such that the events corresponding to the
662~keV $\gamma$-line from $^{137}$Cs have
amplitude larger than $\lsat$=255,
while those from the 32~keV line remain unsaturated.

\begin{figure}
\epsfig{file=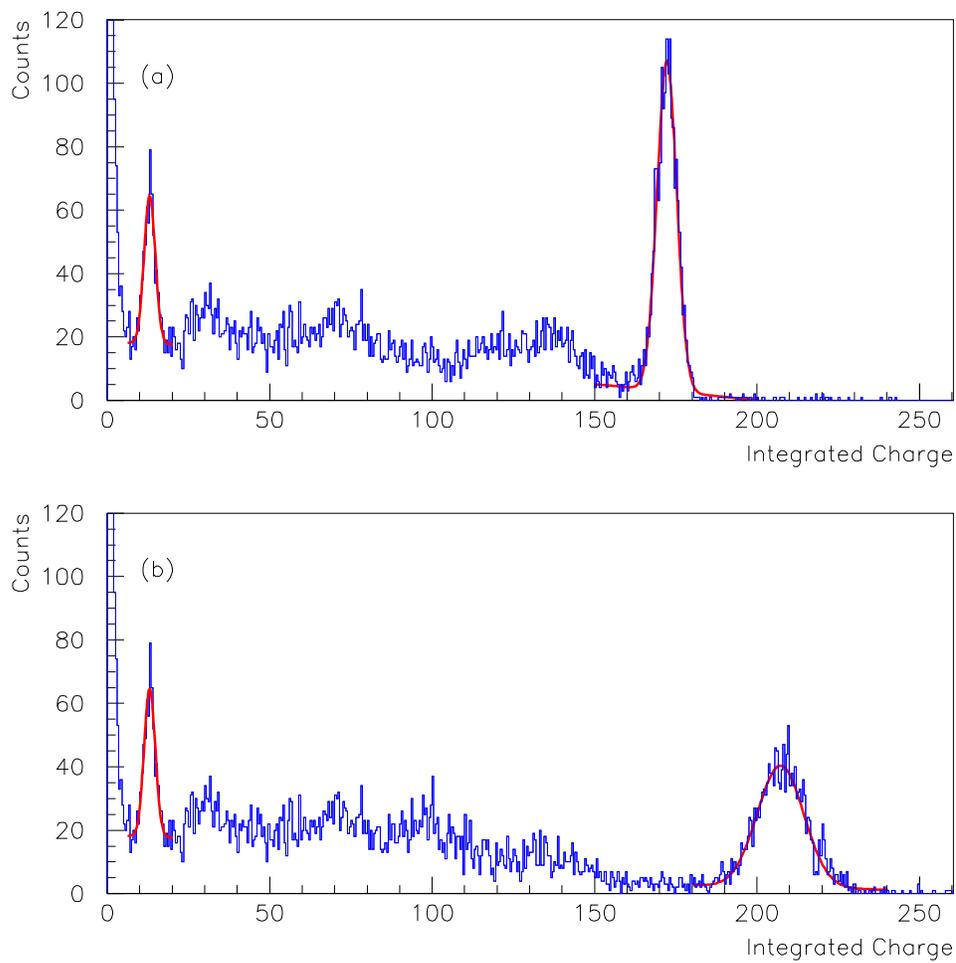,width=15cm}
\caption{
The energy spectra of saturated events (a) before and (b)
after the correction algorithms are applied.
}
\label{hv950}
\end{figure}

\begin{figure}
\epsfig{file=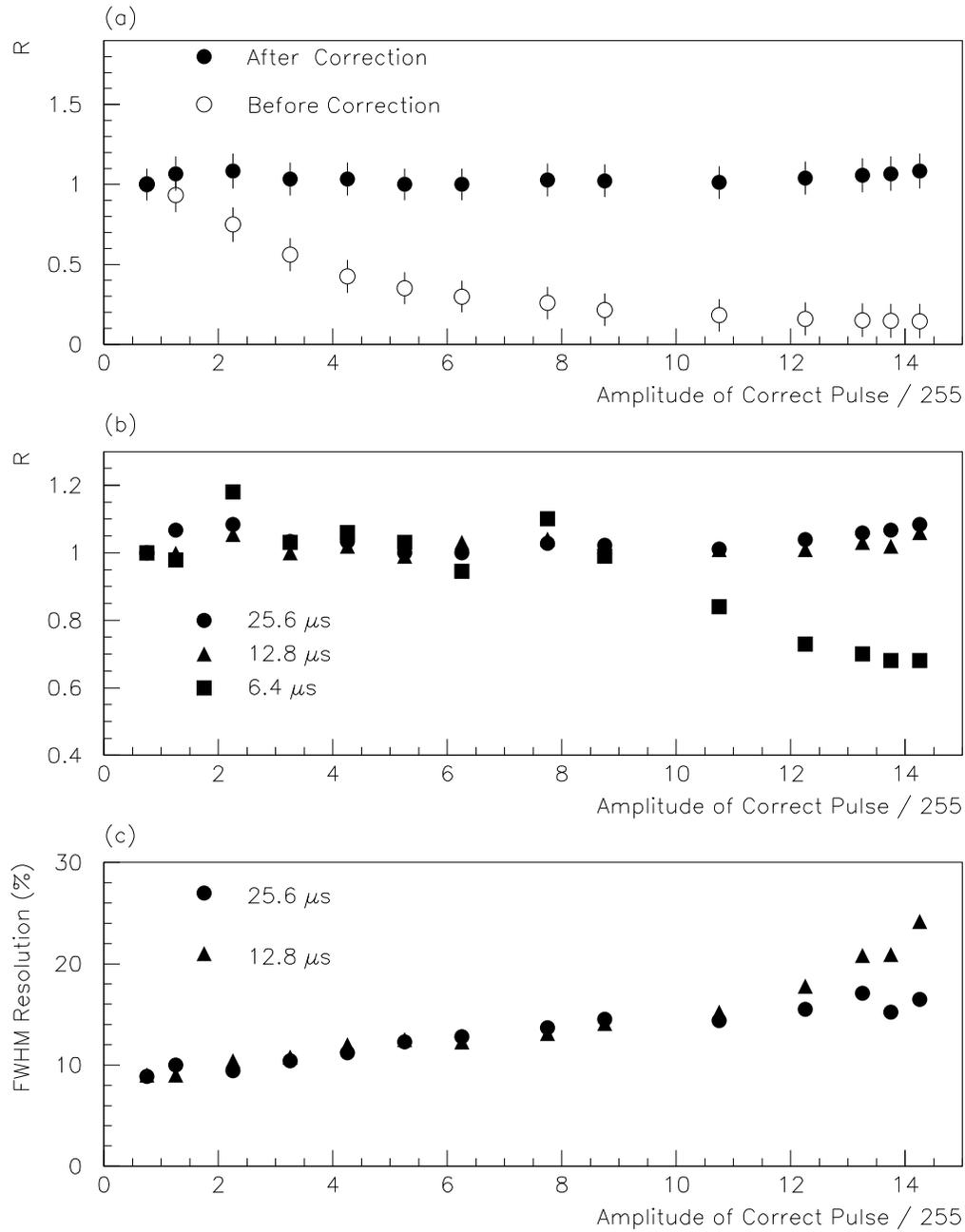,width=15cm}\\
\caption{
(a) The ratios of peak positions 
between the 662~keV and 32~keV $\gamma$-lines before and after
saturated pulse correction; 
(b) the ratios of peak positions at different
time-bin records $\dt2$,
and (c) the energy resolution at 662~keV for
the corrected spectra,
as functions of the expected peak amplitudes.
}
\label{range}
\end{figure}

An example of the energy spectra of the saturated
events, as well as the corrected spectra 
are illustrated in Figure~\ref{hv950}a and 
Figure~\ref{hv950}b, respectively.
The ratio of the
peak positions at 32~keV and 662~keV 
(denoted by $\rm{Q_{32}}$ and $\rm{Q_{662}}$, respectively) 
is measured and compared to the nominal values:
\begin{equation}
\rm{
R ~~ = \frac{ Q_{32} / Q_{662} }{32/662} ~~ .
}
\end{equation}
This can be used as a figure of merit 
to quantify the performance. 
The values of R for saturated and corrected events
as functions of the expected peak amplitudes 
are displayed in Figure~\ref{range}a.
It can be seen that the charge summation correction
methods extend the effective dynamic range of
the FADC by at least a factor of 14, or equivalently,
4 more bits.

\begin{figure}
\epsfig{file=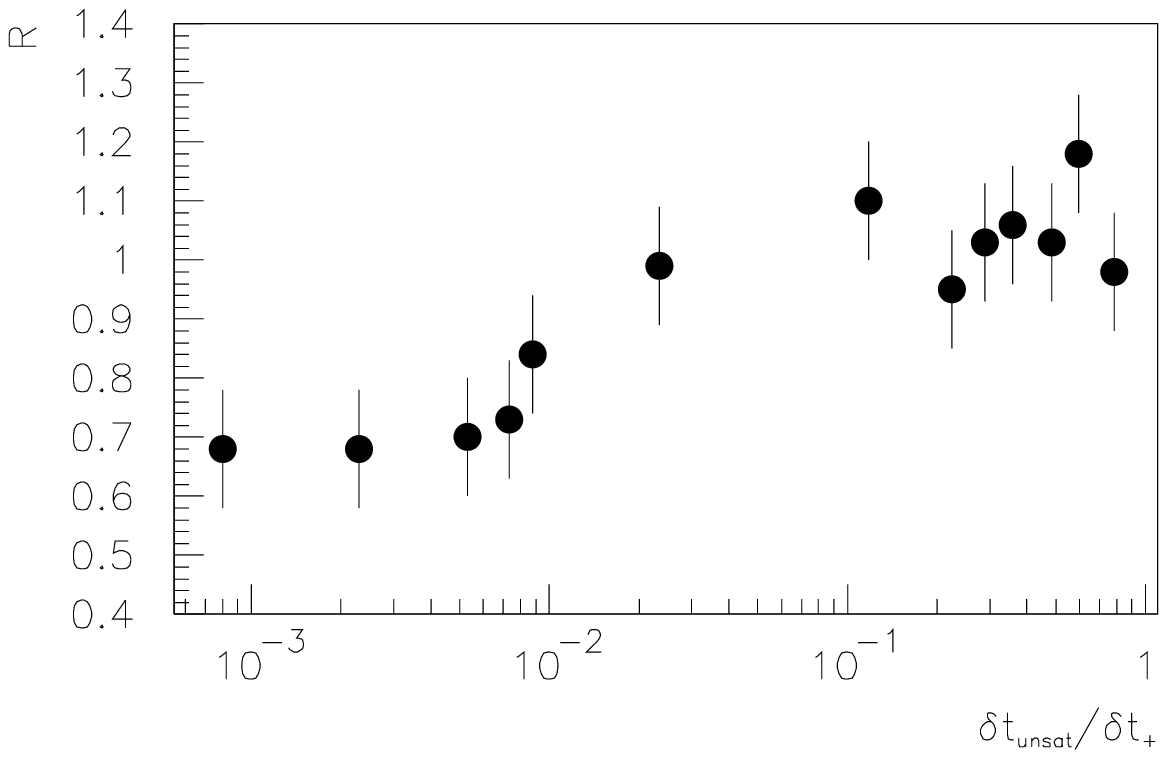,width=15cm}\\
\caption{
Variation of the normalized peak ratio R
with $\dtnosat$/$\dt2$,
where $\dt2$=6.4~$\mu$s and
$\dtnosat$ is the mean unsaturated interval
for the events.
}
\label{dtsat}
\end{figure}

The dependence of performance of the algorithm
on the FADC recording period $\dt2$ is studied.
The data are truncated by software to a shorter
interval and the same
correction procedures are repeated.
The results are displayed in Figure~\ref{range}b
for the variation of R with peak amplitudes
at $\rm{\dt2 = 25.6,~12.8,~6.4 ~ \mu s}$ 
(that is,  512, 256 and 128 time-bins, respectively). 
It can be seen that
the charge reconstruction is no longer valid
for $\rm{\dt2 = 6.4 ~ \mu s}$
at an amplitude of
$>$8  times of $\lsat$.
Alternatively, a plot of  R versus 
$\dtnosat$/$\dt2$ is displayed in
Figure~\ref{dtsat}, where 
$\dtnosat$ is the unsaturated interval, where the
FADC counts are less than $\lsat$.
It can be seen the values of R deviate from
1 only for $\dtnosat$/$\dt2 < 0.02$.
Accordingly, the algorithm is valid so long as the 
the unsaturated interval of the event is more
than 2\% of the FADC recording period.
In comparison, the fitting method as discussed
in Section~\ref{sect::survey} would start to
fail when the saturation interval is more than
twice the value of $\tau_1$, or equivalently, 
the figure-of-merit R deviates from 1 when
$\dtnosat$/$\dt2 < 0.74$.

The dependence of energy resolution 
of the 662~keV line on amplitudes is
shown in Figure~\ref{range}c.
There is no intrinsic bias in the correction,
though the energy resolution gets worse from
around 9\% to 16\% FWHM when the amplitude 
increases by a factor of 14 at $\dt2$=25.6~$\mu$s.
The degradation in energy resolution is larger for 
the shorter FADC recording time.  

\subsubsection{Pulse Shape Discrimination}

\begin{figure}
\epsfig{file=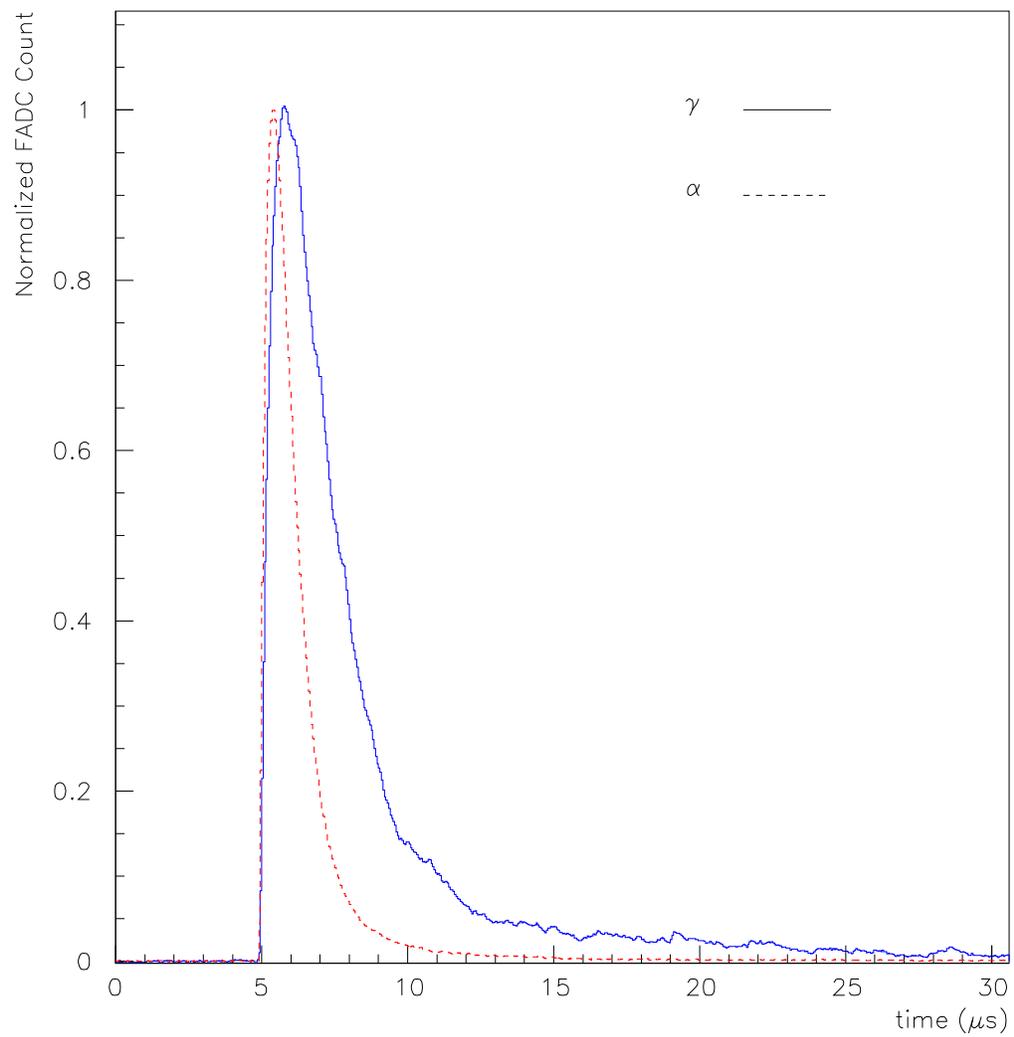,width=15cm}
\caption{
The reference pulse shape for CsI(Tl)
from unsaturated
events due to $\gamma$-rays at 662~keV 
and $\alpha$-particles at 5.4~MeV.
}
\label{pulseshape}
\end{figure}

As depicted in Figure~\ref{pulseshape},
the scintillation light in CsI(Tl) has different decay time
constants for $\gamma$- and $\alpha$-activated events.
The $\alpha$-events decay faster in 
inorganic crystal scintillators $-$ opposite
to the responses in organic 
liquid scintillators~\cite{scinbasic,liqscin}.
Accordingly, the pulse shape information recorded by
the FADC can provide event identification capabilities.

\begin{figure}
\epsfig{file=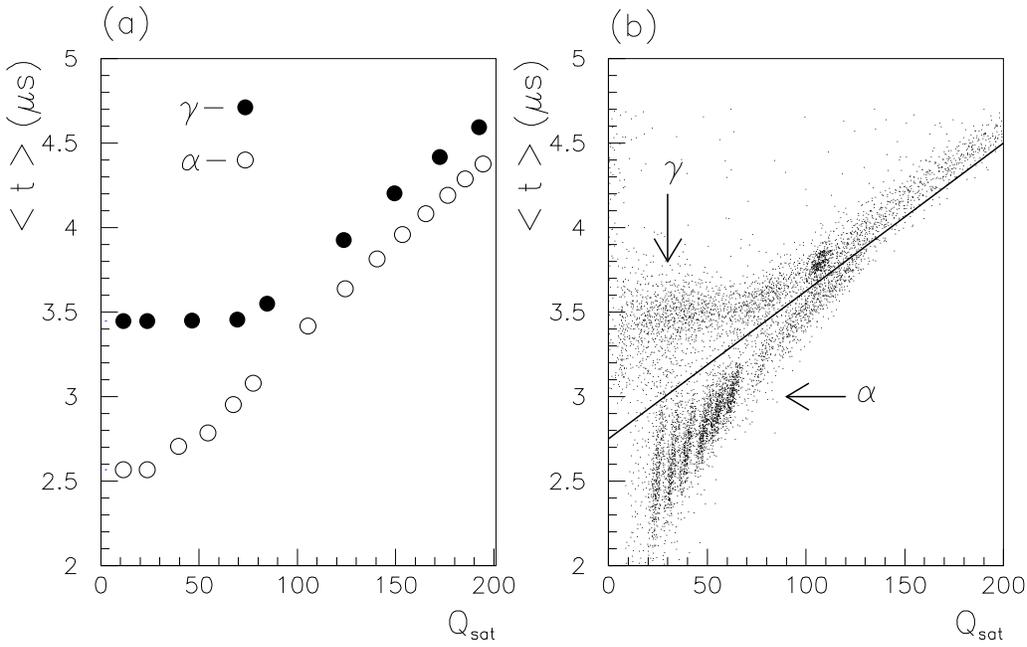,width=15cm}
\caption{
The relationship between $\tbar$
and $\qs$ for 
$\gamma$- and $\alpha$-events:
(a) the average value
and (b) event-by-event scattered plot.
}
\label{tbarqs}
\end{figure}

Identification of $\gamma$/$\alpha$ events
is achieved by studying and comparing
the different responses 
from two observables.
One of them, the saturated charge ($\qs$), has been
adopted in energy re-construction as discussed 
in Section~\ref{sect::partq}. The 
other is the ``mean time'' ($\tbar$) which reflects
the decay profiles, defined as:
\begin{equation}
\rm{
\tbar ~ = ~
\frac{ \sum\limits_{t_i > t_0} ~ [~ t_i - t_0 ~ ] \ast [ ~ Y_i - Y_0 ~ ] }
{ \sum\limits_{t_i > t_0} ~ [~ Y_i - Y_0 ~ ] } ~~~ .
}
\end{equation}
The {\it average} variations of $\tbar$ with $\qs$ 
for both $\gamma$- and $\alpha$-events
can be derived from
the reference pulse shape of Figure~\ref{pulseshape},
and are displayed in Figure~\ref{tbarqs}a.
The scattered plot of $\tbar$ versus
$\qs$ for individual events is
shown in Figure~\ref{tbarqs}b.
It can be seen that $\alpha$/$\gamma$ events
are well separated for $\qs \le 120$.
From Figure~\ref{qtqsamp}, it can be concluded
that $\alpha$/$\gamma$ differentiation
can be achieved for $\alpha$-events with amplitude
up to 20 times of the saturation
level.

\section{Summary}

An algorithm 
to process saturated events in 
Flash Analog-to-Digital Convertor
was developed and studied.
Charge reconstruction and event
identification via pulse shape
analysis are achieved, effectively
extending an 8-bit FADC by
at least 4 more bits.
The algorithm is valid so long as
the unsaturated interval is more 
than 2\% of the FADC recording period.

Although this work originates
from the goals of handling data
from CsI(Tl) crystal scintillators,
the approach should be generic 
and can be applicable to other
detector systems where the entire
pulse shape are recorded while
part of which are saturated.
These studies will facilitate the
specification and design of
large FADC-based data acquisition
systems.

This work was supported by contracts
90-2112-M-001-037 and 91-2112-M-001-036
from the National Science Council, Taiwan,
and 19975050 from the
National Science Foundation, China.

\begin{table}
\begin{center}
\begin{tabular}{|c||c|c|c|c|}
\hline
 & & \multicolumn{2}{|c|}{ } & \\
Event Type & Rise Time & 
\multicolumn{2}{|c|}{Decay Time Constant}
& Ratio (r) \\ \cline{3-4}
 & [$\rm{\tau_0}$ (ns)] & Fast Comp. 
& Slow Comp. & \\ 
 & & 
[$\rm{\tau_1}$ ($\mu$s)] & [$\rm{\tau_2}$ ($\mu$s)] & \\ \hline \hline
 & & & & \\
$\alpha$ & 203$\pm$3 & 0.54$\pm$0.10  & 
2.02$\pm$0.02  & 0.29$\pm$0.02 \\
& & & & \\
$\gamma$ & 261$\pm$2 & 0.87$\pm$0.10  & 
5.20$\pm$0.04  & 0.61$\pm$0.01 \\ 
& & & & \\ \hline
\end{tabular}
\end{center}
\caption{
Fitted rise and decay time constants as
well as the ratio between slow and fast decay components
for $\alpha$ and $\gamma$ events measured by CsI(Tl)
crystal.
}
\label{time}
\end{table}

\clearpage

\clearpage

\clearpage

\clearpage

\clearpage

\clearpage

\clearpage

\clearpage

\clearpage

\clearpage

\end{document}